\documentclass[twocolumn,showpacs,preprintnumbers,amsmath,amssymb]{revtex4}
\usepackage{graphics,epsfig,subfigure}
\usepackage{color}

\begin{document}
\newcommand{\PR}[1]{\ensuremath{\left[#1\right]}} 
\newcommand{\PC}[1]{\ensuremath{\left(#1\right)}} 
\newcommand{\PX}[1]{\ensuremath{\left\lbrace#1\right\rbrace}} 
\newcommand{\BR}[1]{\ensuremath{\left\langle#1\right\vert}} 
\newcommand{\KT}[1]{\ensuremath{\left\vert#1\right\rangle}} 
\newcommand{\MD}[1]{\ensuremath{\left\vert#1\right\vert}} 

\title{Hybrid metric-Palatini brane system}
\author{Qi-Ming Fu\footnote{fuqm12@lzu.edu.cn},
        Li Zhao\footnote{lizhao@lzu.edu.cn, corresponding author},
        Yu-Xiao Liu\footnote{liuyx@lzu.edu.cn, corresponding author}}.
 \affiliation{Institute of Theoretical Physics, Lanzhou University, Lanzhou 730000,
             China}

\begin{abstract}
It is known that the metric and Palatini formalisms of gravity theories have their own interesting features but also suffer from some different drawbacks. Recently, a novel gravity theory called hybrid metric-Palatini gravity was put forward to cure or improve their individual deficiencies. The action of this gravity theory is a hybrid combination of the usual Einstein-Hilbert action and a $f(\mathcal{R})$ term constructed by the Palatini formalism. Interestingly, it seems that the existence of a light and long-range scalar field in this gravity may modify the cosmological and galactic dynamics without conflicting with the laboratory and Solar System tests. In this paper we focus on the tensor perturbation of thick branes in this novel gravity theory. We consider two models as examples, namely, the thick branes constructed by a background scalar field and by pure gravity. The thick branes in both models have no inner structure. However, the graviton zero mode in the first model has inner structure when the parameter in this model is larger than its critical value. We find that the effective four-dimensional gravity can be reproduced on the brane for both models. Moreover, the stability of both brane systems against the tensor perturbation can also be ensured.
\end{abstract}





\maketitle

\section{Introduction}

General relativity is a successful gravitational theory at the scale of the Solar System. However, it does not work well at larger scales. Thus, many modified theories of gravity have been put forward to describe cosmological behaviors such as cosmic acceleration and galactic dynamics \cite{Sotiriou,Felice,Capozziello,Ferreira,Nojiri}. In general, there are three kinds of formalisms for modified gravity theories, namely, the metric formalism, Palatini formalism (matters do not couple with the priori metric-independent connection), and metric-affine formalism (matters couple with the metric and priori metric-independent connection) \cite{Sotiriou}. They all have their own interesting properties and, at the same time, suffer from different drawbacks. Recently, the so-called $C$-theories were proposed in Refs. \cite{Amendola,Koivisto,Koivisto13}, which establishes a bridge between the first and second formalisms in order to find ways to cure or improve their individual deficiencies. In $C$-theories, the Levi-Civita connection $\hat{\Gamma}$ of the metric $\hat{g}_{\mu\nu}$ is conformally related to the spacetime metric $g_{\mu\nu}$, namely, $\hat{g}_{\mu\nu}=\mathcal{C}(\mathcal{R})g_{\mu\nu}$, where $\mathcal{C}$ is an arbitrary function of the Ricci curvature scalar $\mathcal{R}=\mathcal{R}[g,\hat{\Gamma}]=g^{\mu\nu}\mathcal{R}_{\mu\nu}[\hat{\Gamma}]$ only.

Alternatively, another novel modified gravity was presented in Ref.~\cite{Harko}, whose action is a hybrid combination of the usual Einstein-Hilbert action and a $f(\mathcal{R})$ term constructed by the Palatini formalism. It has a dynamically equivalent scalar-tensor representation like the pure metric and pure Palatini cases \cite{Harko,Capozziello1,Capozziello4,Capozziello6}. It also shares properties of both the metric formalism and Palatini formalism like $C$-theories. The new feature of such hybrid gravity is that it predicts the existence of a light long-range scalar field, which can be used to explain the late-time cosmic acceleration~\cite{Harko}.

Considering the characteristics of light and long-range, there is a possibility that this scalar field may modify the cosmological and galactic dynamics \cite{Harko,Capozziello1,Capozziello4,Capozziello6} without conflicting with the laboratory and Solar System tests.
In Ref. \cite{Capozziello4}, the authors analyzed the criteria for obtaining cosmic acceleration and obtained several cosmological solutions, which describe both accelerating and decelerating Universes, depending on the form of the effective scalar potential. The virial theorem was studied in the context of the galaxy cluster, where the mass dispersion relation was modified by a term related to the new scalar field predicted by the hybrid metric-Palatini theory of gravity \cite{Capozziello6}. Stability of the Einstein static Universe was also analyzed in Ref. \cite{static Universe} and a large class of stable solutions were found. In Ref. \cite{wormhole}, the authors considered the possibility that wormhole solutions may be supported by this hybrid metric-Palatini gravity according to the null energy conditions at the throat, and found some specific examples. In Ref. \cite{Capozziello5}, the authors showed that the initial value problem can be well-formulated and well-posed. Moreover, the dynamics of linear perturbation and thermodynamics in hybrid metric-Palatini gravity were also investigated in Refs.~\cite{lima} and \cite{Thermodynamics}, respectively. For a detailed introduction, see the recent review \cite{review}.

On the other hand, it has been extensively considered in the past two decades that our four-dimensional world might be just a brane embedded in a higher-dimensional spacetime. This idea provides new insights into solving some long existing problems, such as the gauge hierarchy and cosmological constant problems \cite{Antoniadis1990,Arkani-Hamed1998,Antoniadis1998,rs1,rs2,Maartens2010}. Dating back to the original Kaluza-Klein (KK) theory, the extra dimension is compacted into a circle with the Planck scale radius. This makes detecting the extra dimension hopeless. While in brane  scenarios, the sizes of extra dimensions can be the order of submillimeter \cite{Antoniadis1998} or infinite \cite{rs2}.

In the Randall-Sundrum-II brane  scenario \cite{rs2}, the thickness of the brane is neglected. In more realistic thick brane models the original singular brane is replaced by a smooth domain wall generated by matter fields. The thick brane models have been extensively studied in the context of higher-dimensional gravity theories \cite{Goldberger1999a,Gremm2000a,Gremm2000b,DeWolfe2000,Csaki2000,Gherghetta2000,Arkani-Hamed2001a,Campos2002,Kobayashi2002,Wang2002,Charmousis2003,Bazeia2004a,Liu2007,
Dzhunushaliev2009,Liu2011,Lu2015}.
The linearization of a brane system is one of the most important issues in brane  models \cite{yangke2012,Cruz2014,Giovannini2001,Bazeia2015a,Kehagias1999,Garriga1999,Bazeia2003,
Cvetic2008,HerreraAguilar2010,Cruz2011,Santos2012,Ahmed2012,Yang2013,Barbosa-Cendejas2013,Higuchi2014,Gu2014}. First, it is a key procedure to investigate stability of the brane solution. Second, to reproduce the effective four-dimensional gravity, we need also to study the linear perturbation of the brane system. Third, the linear perturbation will result in the interaction of matter fields with the KK gravitons, which can be tested by experiments.

In the previous investigations about a brane system, the metric \cite{Yang2013,Cruz2014,Barbosa-Cendejas2013,Higuchi2014,Bazeia2015a} and Palatini formalisms \cite{Gu2014,Bazeia2015b} were individually considered. Therefore, it is interesting to study the properties of a brane system in a gravity theory containing both formalisms. For example,
how does the hybrid of the two formalisms affect the properties of the brane solutions and stability of the linear perturbation? This motivates us to investigate the hybrid metric-Palatini brane system.
In this paper, inspired by the scalar-tensor representation of hybrid metric-Palatini gravity, we will consider two models: the thick branes constructed by a background scalar field (model A) and by pure gravitational system (model B) in hybrid metric-Palatini gravity. In Refs. \cite{Dzhunushaliev2010,Lu1202,zhong2015,Parry,Bronnikov2007,Deruelle2008,Balcerzak2008,Herrera-Aguilar2010}, some brane  models have been constructed for pure gravitational systems without matter fields. This scenario is the same as producing an expanding universe from $f(R)$ gravity without introducing an extra scalar field (inflation without inflaton). In order to study the issues of stability of the tensor perturbation and localization of the graviton zero mode, we will investigate the linearization of these two brane  models.

In this work, we focus on brane  in hybrid metric-Palatini gravity. Therefore,
in Sec. \ref{metricpalatini} we briefly introduce the hybrid metric-Palatini model and find thick brane  solutions
for model A and model B. Stability of the brane system and localization of the graviton zero mode are analyzed
in Sec. \ref{localization}. Section \ref{conclusion} comes with the conclusion.

\section{The Hybrid metric-Palatini brane models and solutions}\label{metricpalatini}
Now, let us start with the action of the five-dimensional brane  model in hybrid metric-Palatini gravity \cite{Harko}.
\begin{equation}{\label{action1}}
 S = \frac{1}{2\kappa^2} \int d^5 x \sqrt{-g}\Big[R + f(\mathcal{R})\Big] + S_{m}(g,\chi),
\end{equation}
where $\kappa^2 = 8 \pi G_5$ with $G_5$ the five-dimensional Newtonian gravitational constant and we have set $c=1$. $S_m$
is the standard matter action, $R=g^{MN}R_{MN}$ is the Einstein-Hilbert Ricci scalar constructed by the metric, and $\mathcal{R} = g^{MN} \mathcal{R}_{MN}$ is the Palatini curvature, where $\mathcal{R}_{MN}$ is defined in terms of a torsionless independent connection, $\hat{\Gamma}$, as
\begin{eqnarray}
\mathcal{R}_{MN} \!\!\equiv\!\! \PC{\hat{\Gamma}^{P}_{MN, P} - \hat{\Gamma}^{P}_{M P,N} + \hat{\Gamma}^{P}_{P Q} \hat{\Gamma}^{Q}_{MN} - \hat{\Gamma}^{P}_{M Q} \hat{\Gamma}^{Q}_{P N}}.
\end{eqnarray}

Introducing an auxiliary scalar field $\phi$, the action (\ref{action1}) can be deformed as
\begin{equation}{\label{action2}}
 S = \frac{1}{2 \kappa^2} \int d^5 x \sqrt{-g} \Big[R + \phi \mathcal{R} - V_1(\phi)\Big] + S_{m},
\end{equation}
\noindent where
\begin{eqnarray}~\label{re}
 \phi \equiv F(\mathcal{R}) = \frac{d f(\mathcal{R})}{d \mathcal{R}}, ~~~ V_1(\phi) \equiv \mathcal{R} F(\mathcal{R}) - f(\mathcal{R}).
\end{eqnarray}

The field equations can be obtained by varying the action (\ref{action2}) with respect to the metric $g_{MN}$, the scalar field $\phi$, and the independent connection $\hat{\Gamma}^{P}_{MN}$:
\begin{subequations}
\begin{eqnarray}
R_{MN} + \phi \mathcal{R}_{MN} - \frac{1}{2}\PC{R + \phi \mathcal{R} - V_1}g_{MN} \!\!&=&\!\! \kappa^2 T_{MN},~~~~~~\label{eom1}\\
\mathcal{R} - V_{1\phi} &=& 0,~\label{eom2}\\
\hat{\nabla}_{P} \PC{\sqrt{-g} \phi g^{MN}} &=& 0,~\label{eom3}
\end{eqnarray}
\end{subequations}
where $V_{1\phi}\equiv\frac{dV_1(\phi)}{d\phi}$, the matter stress-energy tensor is defined as usual $T_{MN}\equiv -\frac{2}{\sqrt{-g}}\frac{\delta(\sqrt{-g}\mathcal{L}_m)}{\delta g^{MN}}$, and $\hat{\nabla}_{P}$ is compatible with the connection $\hat{\Gamma}^{P}_{MN}$.

The solution of Eq.~(\ref{eom3}) implies that the independent connection is the Levi-Civita connection of a metric $q_{MN}=\phi^{2/3}g_{MN}$. Then, the relation between $\mathcal{R}_{MN}$ and $R_{MN}$ is
\begin{eqnarray}{\label{riccirelation}}
 \mathcal{R}_{MN} &=& R_{MN} + \frac{4}{3\phi^2} \partial_M\phi\partial_N\phi-\frac{1}{\phi}\Big(\nabla_M\nabla_N\phi \nonumber\\
 &+&\frac{1}{3}g_{MN}\Box\phi\Big),
\end{eqnarray}
where $\Box\equiv g^{KL}\nabla_K\nabla_L$. Using the relation (\ref{riccirelation}), one can obtain a scalar-tensor representation \cite{Harko,Capozziello1,Capozziello4}:
\begin{eqnarray}~\label{action3}
S&=& \frac{1}{2\kappa^2}  \int d^5 x \sqrt{-g} \Big[(1+\phi)R+\frac{4}{3\phi}\partial^{K}\phi\partial_{K}\phi \nonumber\\
 &-& V_1(\phi)\Big] + S_m.
\end{eqnarray}

Now, it is clear that the free choice of the form of the $f(\mathcal{R})$ is transformed to the potential $V_1(\phi)$ of a scalar profile $\phi$. Inspired by the scalar-tensor representation, we consider two models:  model A for the brane constructed by a matter scalar field $\chi$ and
 model B for the brane constructed by the pure gravity without any matter field.

\subsection{Model A: with matter}

The action of the matter part with a scalar field is
\begin{eqnarray}
S_m=\int d^5 x \sqrt{-g}\Big[-\frac{1}{2}g^{MN}\partial_{M}\chi\partial_{N}\chi-V_2(\chi)\Big]. \label{matter}
\end{eqnarray}
Substituting Eq.~(\ref{riccirelation}) and Eq.~(\ref{eom2}) in Eq.~(\ref{eom1}), the gravitational field equation can be written as
\begin{eqnarray}
\!\!\!&&(1+\phi)R_{MN}+\frac{4}{3\phi}\partial_M\phi\partial_N\phi-\Big(\nabla_M\nabla_N\phi-g_{MN}\Box\phi\Big) \nonumber\\
\!\!\!&&-\frac{1}{2}g_{MN}\Big[(1+\phi)R+\frac{4}{3\phi}\partial^K\phi\partial_K\phi-V_1(\phi)\Big]=\kappa^2 T_{MN}. \nonumber\\ \label{eom11}
\end{eqnarray}
Considering Eq.~(\ref{eom2}) and the trace of Eq.~(\ref{riccirelation}), one finds that  the scalar field $\phi$ is governed by the second-order evolution equation
\begin{eqnarray}~\label{eom22}
  8\phi \Box\phi
   -4\partial_K\phi\partial^K\phi
   -\phi^2\big[5V_1(\phi)-3(1+\phi)V_{1\phi}\big] \nonumber\\
   +2\phi^2\kappa^2 T=0.
\end{eqnarray}

Our discussion will be limited to the static flat brane  scenario, for which the metric is given by
\begin{eqnarray}~\label{metric}
ds^2=\text{e}^{2A(y)}\eta_{\mu\nu}dx^{\mu}dx^{\nu}+dy^2,
\end{eqnarray}
with $y$  the extra dimension. Meanwhile, the scalar field, $\phi=\phi(y)$, is independent of the brane coordinates.
For the system (\ref{eom11})-(\ref{metric}), the Einstein field equations and equation of motion of the scalar field $\phi$ are read as
\begin{subequations}~\label{eom5}
\begin{eqnarray}
3 (A''+4A'^2)(1+\phi)+7 A' \phi ' +\text{V}_1\nonumber\\
   +2 \kappa ^2 \text{V}_2+\phi '' &=& 0,~\label{eom110}\\
 12 (A''+A'^2)(1+\phi)+4 A' \phi ' +\text{V}_1 +4 \phi ''\nonumber\\
   +2 \kappa ^2 \text{V}_2 +3 \kappa ^2 \chi '^2
   -{4 \phi '^2}/{\phi } &=& 0,~\label{eom115}\\
\!\!\!32 A' \phi'^2 \phi-\phi^2  \left(5 \phi'\text{V}_1-{3 (1+\phi) \text{V}_1'}\right) -4\phi '^3\nonumber\\
- \kappa ^2 \phi'\phi^2  \left(10 \text{V}_2+3 \chi '^2\right)
 + 8 \phi ''\phi'\phi &=& 0,~~~~~~\label{eom222}
\end{eqnarray}  \label{EoMsModelA}
\end{subequations}
where a prime stands for the derivative with respect to the extra-dimensional coordinate $y$.

The equation of motion of the matter field is described by the following equation:
\begin{eqnarray}~\label{mattereom}
4A'\chi'+\chi''=\frac{dV_2(\chi)}{d\chi}.
\end{eqnarray}
Equations (\ref{eom5})-(\ref{mattereom}) describe the whole system. There are five variables, namely, $A(y),~ \phi(y),~\chi(y),~V_1(\phi)$, and $V_2(\chi)$.
However, there are only three independent equations. So, one needs assume two conditions to solve this system.

\begin{figure*}[htb]
\begin{center}
\subfigure[$\text{e}^{2A(y)}$]  {\label{warpfactor1}
\includegraphics[width=5cm]{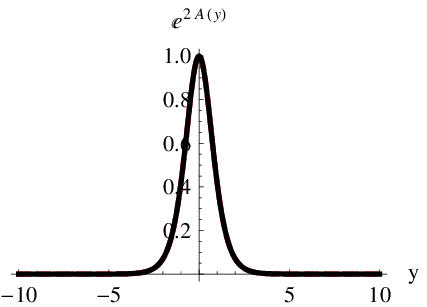}}
\subfigure[$\phi(y)$]  {\label{phi1}
\includegraphics[width=5cm]{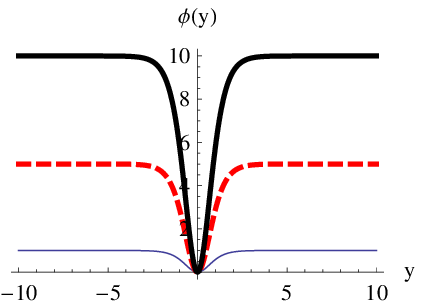}}
\subfigure[$\chi(y)$]  {\label{chi}
\includegraphics[width=5cm]{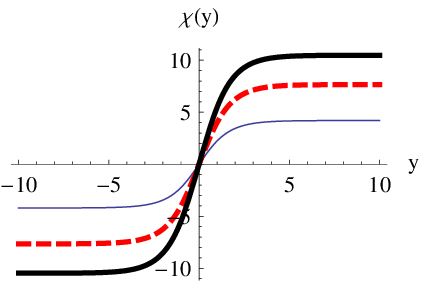}}
\subfigure[$V_1(y)$]  {\label{Vphi1}
\includegraphics[width=5cm]{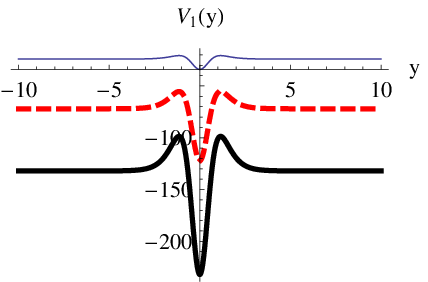}}
\subfigure[$V_2(y)$]  {\label{Vchi}
\includegraphics[width=5cm]{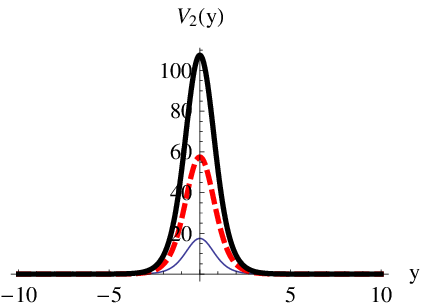}}
\subfigure[$\rho(y)$]  {\label{rho1}
\includegraphics[width=5cm]{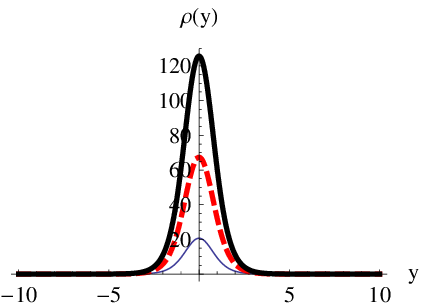}}
\end{center}
\caption{Plots of the brane solution (\ref{BraneSolutionModelAa})-(\ref{BraneSolutionModelAb}) and energy density for model A. The parameters are set to $b=1$, $k=1$, $a=1$ for thin lines, $a=5$ for red dashed thick lines, $a=10$ for black thick lines.}
\label{A}
\end{figure*}

In this model, we consider the following configuration for the scalar field $\phi(y)$ and warp factor $A(y)$:
\begin{subequations}
\begin{eqnarray}
\phi(y) &=& a~\text{tanh}^2(ky),  \\
A(y)   &=&  b~\text{ln}[\text{sech}(ky)].
\end{eqnarray}\label{BraneSolutionModelAa}
\end{subequations}
In order to avoid the ghost problem, we should ensure the positive definiteness of the coefficient of $R$ in the action (\ref{action3}), so we should take $a>0$.
Now it can be checked that the system supports the following solutions:
\begin{subequations}
\begin{eqnarray}
\chi(y)\!\!\!&=&\!\!\!
       \tanh (k y){\sqrt{\frac{1}{6} \Big(3 (5 a+3) \cosh (2 k y)+5 a+9\Big)}}\nonumber\\
      \!\!&+&\!\! i \sqrt{\frac{10 a+9}{3}}
         \Big[  E\Big(i k y,\frac{15 a+9}{10 a+9}\Big) \nonumber\\
               &-& F\Big(i k y,\frac{15 a+9}{10 a+9}\Big)
          \Big],~~~~~~~~~ \\
V_1(y) &=& -\frac{1}{2} k^2 \text{sech}^4(k y) \Big[(12-8 a) \cosh (2 k y) \nonumber\\
       &+& 3 (a+1) \cosh (4 k y)+49 a+9\Big],\\
V_2(y) \!\!&=&\!\! \frac{5}{2} k^2 \text{sech}^2(k y) \Big[-a ~\text{sech}^2(k y)+5 a+3\Big],
\end{eqnarray}\label{BraneSolutionModelAb}
\end{subequations}
where we have chosen $b=1$ and $\kappa=1$ for simplicity, and the functions $E$ and $F$ are two kinds of elliptic integrals. The shapes of this brane solution and the energy density are shown in Fig.~\ref{A}. Obviously, the energy density peaks at the origin of the extra dimension, which represents a single brane. It is not difficult to analyze the structure of the five-dimensional spacetime at $y=\pm\infty$, where the curvature $R= -20k^2 < 0$. This means that the spacetime is asymptotic AdS.


We can also obtain the expression of $f(\mathcal{R})$ from Eq.~(\ref{re})
\begin{eqnarray}
f(\mathcal{R})=\frac{2a}{3}\mathcal{R} - \frac{a}{120 k^2}\mathcal{R}^2 + \frac{26a k^2}{3} + 12 k^2.
\end{eqnarray}
Then, the complete Lagrangian for gravity can be expressed as
\begin{eqnarray}
\mathcal{L}=R+\frac{2a}{3}\mathcal{R} - \frac{a}{120 k^2}\mathcal{R}^2 + \frac{26a k^2}{3} + 12 k^2.
\end{eqnarray}

\subsection{Model B: without matter}

We can also construct a brane from the scalar profile $\phi(y)$ without introducing the matter field $\chi(y)$. Then we can obtain the field equations of the whole system just by omitting the terms about the matter field $\chi(y)$ from Eqs. (\ref{EoMsModelA}):
\begin{subequations}
\begin{eqnarray}
3 (1+\phi) (A''+4  A'^2)+7 A' \phi '+\text{V}_1(\phi)+\phi '' &=& 0,~\label{eom120}\\
12 \phi (1+\phi)  (A''+A'^2) + 4 A' \phi' \phi+ \phi\text{V}_1(\phi) \nonumber\\
          + 4 \phi \phi '' - 4 \phi '^2 &=& 0,~~~~~~~\label{eom125}\\
8\phi'' +32 A'\phi'-5\phi\text{V}_1(\phi)+3\phi(1+\phi)\text{V}_{1\phi} &=& 0.
\end{eqnarray} \label{EoMsModelB}
\end{subequations}
Now, there are only three variables, namely, $A(y),~\phi(y)$, and $V_1(\phi)$, but only two independent equations.
So, we just need one condition to solve this system.

Subtracting (\ref{eom120}) from (\ref{eom125}) yields
\begin{eqnarray}
9\phi(1+\phi)A''+3\phi\phi''-3{\phi}A'\phi'-4\phi'^2=0.~\label{eoms}
\end{eqnarray}
It is easy to check that this equation yields a thin brane solution, i.e., $\phi(y)=\phi_1$ and $A(y)=-\alpha|y|$, where both $\phi_1$ and $\alpha$ are constants. In this paper, we mainly focus on thick brane solution, so we suppose
\begin{eqnarray}
A(y)=b~\text{ln}[\text{sech}(ky)], \label{warp2}
\end{eqnarray}
where $b$ is a positive parameter in order to localize the graviton zero mode on the brane (see Sec. \ref{localization}).
In order to keep the $Z_2$ symmetry of the extra dimension, we only look for even function solution for the scalar $\phi(y)$. Thus, the initial condition for $\phi(y)$ can be assumed as
\begin{eqnarray}
\phi(0)=\phi_0, \quad\quad  \phi'(0)=0,~\label{InitialCondition}
\end{eqnarray}
from which and Eq.~(\ref{eoms}), we can get
\begin{eqnarray}~\label{phiof0}
\phi''(0)=3(1+\phi_0)k^2 b.
\end{eqnarray}
In order to ensure the positive definiteness of the coefficient of $R$ in the action (\ref{action3}), we require $1+\phi(y)>0$, from which one has $1+\phi_0>0$ and so $\phi''(0)>0$.

\begin{figure}[htb]
\begin{center}
\subfigure[$\phi(y)$]  {\label{Valueofphi01}
\includegraphics[width=4cm]{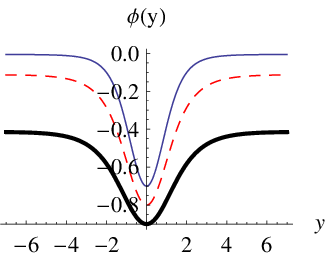}}
\subfigure[$V_1(y)$]  {\label{Vphi2}
\includegraphics[width=4cm]{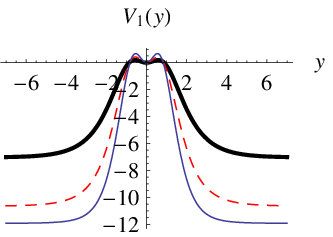}}
\end{center}
\caption{Plots of the scalar profile $\phi(y)$ and scalar potential $V_1(y)$ for model B. The parameters are set to $b=1$, $k=1$. The black thick, red dashed thin, blue thin lines correspond to $\phi_0=-0.9$, $-0.8$, $-0.7$ respectively.}
\label{Bofphi}
\end{figure}

Considering the asymptotic behaviour of the warp factor $A(y\rightarrow \pm\infty)\rightarrow -bk|y|$,
one can obtain the asymptotic behavior of the scalar profile from Eq.~(\ref{eoms}):
\begin{eqnarray}~\label{asym}
\phi(y\rightarrow \pm\infty)\rightarrow\frac{c_1}{(\text{e}^{c_2}+\text{e}^{-bk|y|})^3}\rightarrow c_1,
\end{eqnarray}
where $c_1$ and $c_2$ are integral constants related to $\phi(0)$ and $\phi'(0)$.
The numerical solutions of the scalar field and scalar potential $V_1(y)$ are plotted in Fig. \ref{Bofphi}, from which one can see that $c_1$ increases with $\phi(0)$.
Note that the even parity of the scalar $\phi(y)$ here would have different influence on the localization of fermions \cite{XuZengguang} from the case of odd scalar kink solutions in other brane models \cite{Cvetic1997,Gremm2000a}.

\section{Localization of gravity}\label{localization}

Since we already have two brane models (model A and model B), we will consider stability of these two models under the tensor perturbation of the spacetime metric and the localization of the graviton zero mode, which are two important issues in brane  theory. Generally speaking, the four-dimensional massless graviton should be localized on the brane in order to reproduce the familiar four-dimensional Newtonian potential. We will analyze these issues in the following context.

Since the scalar, vector, and tensor fluctuations are decoupled from each other, we can write the spacetime metric under the tensor fluctuation as
\begin{eqnarray}
ds^2=e^{2A(y)}(\eta_{\mu\nu}+h_{\mu\nu})dx^{\mu}dx^{\nu}+dy^2,
\end{eqnarray}
where $h_{\mu\nu}$ represents the tensor fluctuation and it is transverse-traceless, i.e., $\eta^{\mu\beta}\partial_{\beta}h_{\mu\nu}=0$ and $h\equiv \eta^{\mu\nu}h_{\mu\nu}=0$. The field equation of the tensor perturbation reads
\begin{eqnarray}~\label{flueq1}
h''_{\mu\nu}+\Big(4A'+\frac{\phi'}{1+\phi}\Big)h'_{\mu\nu}+\text{e}^{-2A}\Box^{(4)}h_{\mu\nu}=0,
\end{eqnarray}
where $\Box^{(4)}=\eta^{\mu\nu}\partial_{\mu}\partial_{\nu}$ stands for the four-dimensional D'Alembertian operator.
By making a coordinate transformation $dy=\text{e}^{A}dz$, Eq. (\ref{flueq1}) can be rewritten as
\begin{eqnarray}~\label{flueq2}
\partial_z^2 h_{\mu\nu}+\Big(3 \partial_z A+\frac{\partial_z\phi}{1+\phi}\Big) \partial_z h_{\mu\nu}+\Box^{(4)}h_{\mu\nu}=0.
\end{eqnarray}

After making the KK decomposition $h_{\mu\nu}=\varepsilon_{\mu\nu}(x)f(z)H(z)$, we can get the following two equations:
\begin{subequations}
\begin{eqnarray}
&&\Box^{(4)}\varepsilon_{\mu\nu}(x) = m^2 \varepsilon_{\mu\nu}(x),~\label{KG}\\
\!\!\!\!\!\!\!&-&\!\!\partial_{z}^2 H(z)
  - \Big( 3\partial_z A+\partial_z \ln \big(f^2(1+\phi)\big) \Big)\partial_z H(z) \nonumber\\
&-&\!\! \Big(\frac{\partial_{z}^2 f}{f} + 3 \frac{\partial_z A~\partial_z f}{f} + \frac{\partial_z \phi}{1+\phi}\frac{\partial_z f}{f}\Big) H(z) = m^2 H(z).\nonumber\\
\label{ls}
\end{eqnarray}
\end{subequations}
where Eq.~(\ref{KG}) is the Klein-Gordon equation for the four-dimensional massless ($m=0$) or massive ($m\neq 0$) graviton.
In order to get a Schr$\ddot{\text{o}}$dinger-like equation of the KK mode $H(z)$, its first-order derivation  should be vanishing. Thus, the function $f(z)$ can be solved from $ 3\partial_z A+\partial_z \ln \big(f^2(1+\phi)\big) =0$ as
\begin{eqnarray}
f(z)=\frac{\text{e}^{-{3A}/{2}}}{\sqrt{1+\phi}}.
\end{eqnarray}
Then, Eq.~(\ref{ls}) can be rewritten as
\begin{eqnarray}
\big(-\partial_z^2+U(z)\big)H(z) &=& m^2 H(z),~\label{SchrodingerEq}
\end{eqnarray}
where the effective potential for the KK mode is given by
\begin{eqnarray}
U(z) 
 &=& 2\frac{(\partial_{z}f)^2}{f^2}-\frac{\partial_{z}^2 f}{f} \nonumber \\
 &=& \frac{3 }{2}\partial_{z}^2A+\frac{9}{4} (\partial_{z}A)^2
     -\frac{ (\partial_{z}\phi)^2}{4 (1+\phi)^2} \nonumber \\
 &&  +\frac{3 \partial_{z} A ~\partial_{z} \phi  + \partial_{z}^2\phi}{2(1+ \phi)}, \label{EffectivePotentialU}
\end{eqnarray}
which can be rewritten in coordinate $y$ as
\begin{eqnarray}
U(z(y))
 &=& e^{2 A} \Big(\frac{3}{2} A''+\frac{15}{4} A'^2-\frac{\phi '^2}{4 (\phi+1)^2} \nonumber\\
 &+&\frac{4 A' \phi '+\phi ''}{2 (\phi +1)}\Big). \label{EffectivePotentialUy}
\end{eqnarray}
Equation (\ref{SchrodingerEq}) is the equation of motion for the KK mode $H(z)$, and it can be factorized as the supersymmetric form $L^\dag L H(z)=m^2 H(z)$ with $L=(\frac{d}{dz}+\frac{\partial_z f}{f})$ and $L^{\dag}=(-\frac{d}{dz}+\frac{\partial_z f}{f})$.
The Hermitian and positive definite of the operator $L^{\dag}L$ ensure that $m^2\geqslant 0$.
Thus, there is no tachyonic KK mode.

By setting $m=0$ in Eq.~(\ref{SchrodingerEq}), the graviton zero mode can be solved as
\begin{eqnarray}
H_0(z)=N_0 f^{-1}(z)=N_0 \sqrt{1+\phi}~\text{e}^{{3A}/{2}},
\end{eqnarray}
where $N_0$ is a normalization constant. The normalization of the zero mode is expressed as
\begin{eqnarray}
\int \!H_0^2 dz=\int \!H_0^2 ~{e^{-A}} {dy}=N_0^2\int \!\big(1+\phi\big)\text{e}^{2A}dy=1.~~\label{normalization}
\end{eqnarray}

\begin{figure}[htb]
\begin{center}
\subfigure[$U(y)$]  {\label{U1}
\includegraphics[width=4cm]{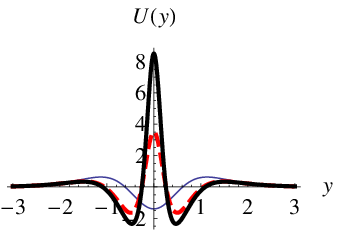}}
\subfigure[$H_0(y)$]  {\label{H01}
\includegraphics[width=4cm]{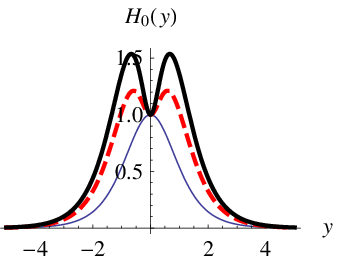}}
\end{center}
\caption{Plots of the effective potential $U(y)$ of the gravitational fluctuation and the graviton zero mode $H(y)$ for model A. The parameters are set to $b=1$, $k=1$, $\kappa=1$, and $a=0.1$ for thin lines, $a=5$ for red dashed thick lines, $a=10$ for black thick lines.}
\label{c}
\end{figure}

For model A with the other parameters set to $b=1$, $k=1$, and $\kappa=1$, $N_0$ can be calculated as $N_0=\sqrt{\frac{3}{6+2a}}$. For arbitrary positive parameters, it can be shown that the integral in (\ref{normalization}) is finite. So, the graviton zero mode in model A can be localized on the brane. Figure \ref{c} shows the shapes of the effective potential of the gravitational fluctuation and the nonnormalized graviton zero mode. It can be seen that the shape of the effective potential changes from volcano-like well to double well with increasing $a$. From Eq.~(\ref{EffectivePotentialUy}), we can obtain $U''(0)=-4 a^2-18 a+{27}/{2}$. It is obvious that the shape of the effective potential is volcano-like for $U''(0)>0$ and double well for $U''(0)<0$. The critical value of the parameter $a$ is $a_{\text{c}}=\frac{3}{4} \left(\sqrt{15}-3\right)$ since we only need positive $a$. Thus, the effective potential has a single well and a double well for $0<a<a_{\text{c}}$ and $a>a_{\text{c}}$, respectively.

It can also be seen that the graviton zero mode is localized gradually far away from the origin of the extra dimension (see Fig.~\ref{H01}) because the shape of the effective potential changes from volcano like to double well, see Fig.~\ref{U1}. This character does not mean a double brane but a single brane because the energy density still peaks at the origin of the extra dimension (see Fig.~\ref{rho1}). Therefore, even although the brane has no internal structure, the effective potential has a double well and the graviton zero mode has a split for large $a$.  This is a new result of this model that different from the previous ones in the literatures.

\begin{figure}[htb]
\begin{center}
\subfigure[$U(y)$]  {\label{U2}
\includegraphics[width=4cm]{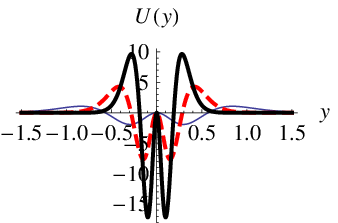}}
\subfigure[$H_0(y)$]  {\label{H02}
\includegraphics[width=4cm]{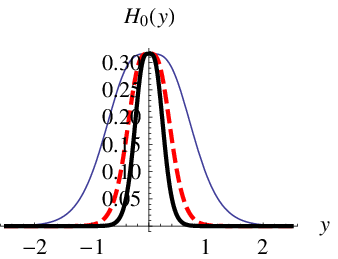}}
\end{center}
\caption{Plots of the effective potential of the gravitational fluctuation and the graviton zero mode of model B. The parameters are set to $b=3$, $\phi_0=-0.9$, and $k=1$ for thin lines, $k=2$ for red dashed thick lines, $k=3$ for black thick lines.}
\label{C}
\end{figure}

Figure \ref{C} shows the shapes of the effective potential and graviton zero mode in model B.
From Eqs.~(\ref{warp2}), (\ref{InitialCondition}), (\ref{phiof0}), and (\ref{EffectivePotentialUy}), one can get \begin{eqnarray}
  U(0)=\frac{\phi ''(0)}{2(\phi_0+1)}-\frac{3}{2} b k^2=\frac{3}{2} b k^2-\frac{3}{2} b k^2=0. \nonumber
\end{eqnarray}
Thus, the shape of the effective potential of the
gravitational fluctuation is always a double well for any positive parameters $b$ and $k$.
From the asymptotic behaviour of the warp factor $A(y\rightarrow \pm \infty) \rightarrow -bk|y|$ and scalar profile $\phi(y\rightarrow \pm \infty)\rightarrow c_1$,
it is easy to check that the corresponding graviton zero mode for model B can also be normalizable:
$\int H_0^2(z) dz<\infty$. Therefore, the graviton zero mode can be localized on the brane.

So, we can conclude that the brane can be constructed by the background scalar field or by pure gravity,
and the Newtonian potential on the brane can be reproduced on both models since the graviton zero mode can be localized on the brane.

\section{Conclusions and Discussion}\label{conclusion}

In this paper, we investigated two thick brane  models (model A and model B) in hybrid metric-Palatini gravity. The brane in model A was constructed by a background scalar field $\chi$. This brane system can be solved analytically. On the other hand, inspired by the scalar-tensor representation of this gravity, we considered the possibility of the thick brane constructed by pure gravity, which is called model B. We obtained a set of numerical solutions for the brane system in this model. Then, we derived the field equation of the tensor perturbation (\ref{flueq1}). After the KK decomposition, we obtained a Schr$\ddot{\text{o}}$dinger-like equation of the KK modes $H(z)$, which is the equation of motion of the graviton along the extra dimension.
This equation can be factorized as a supersymmetric form, which ensures the stability of the brane system.
Furthermore, we also gave the condition that avoids the ghost gravitons.

In order to produce the four-dimensional Newtonian potential, we analyzed the graviton zero modes in both models. The graviton zero mode in model A splits from one peak to two peaks with the parameter $a$ increasing; however, the brane does not split. This means that the graviton zero mode is localized gradually far away from the origin of the extra dimension with increasing $a$. The reason is that the shape of the effective potential of the gravitational fluctuation changes from volcano like ($0<a<a_{\text{c}}$) to double well ($a>a_{\text{c}}$).
The graviton zero mode in model B is localized around the origin of the extra dimension and becomes thinner with the parameter $k$ increasing. The shape of the effective potential of the gravitational fluctuation is always a double well. The graviton zero modes in both models are localized on the branes. So, we can obtain the familiar four-dimensional Newtonian potential for both models.

The localization of various matter fields on the branes in the two models (especially the one constructed by pure gravity) is an important and interesting question. It leaves for our future work.

\acknowledgments{
This work was supported by the National Natural Science Foundation of China (Grants No. 10905027, No. 11375075, and No. 11522541),
and the Fundamental Research Funds for the Central Universities (Grant No. lzujbky-2015-jl01).}

\end{document}